\documentclass[11pt]{article}
\usepackage{graphicx}
\usepackage{amssymb}
\usepackage{amscd}
\usepackage{mathrsfs}
\usepackage{longtable}
\usepackage{lscape}
\usepackage{amsthm}
\usepackage{amsfonts}
\usepackage{amsmath}
\usepackage{bbm}
\usepackage{float}
\usepackage{url}
\usepackage[polish,english]{babel} 
\usepackage[utf8]{inputenc} 
\usepackage[T1]{fontenc}    
\usepackage{hyperref}
\usepackage[margin=0.5cm,font=footnotesize]{caption}
\usepackage{lineno}
\usepackage[numbers]{natbib} 
\usepackage{appendix}
\usepackage{subfig}

\title{\textbf{Algorithmic Idealism II: Reassessment of Competing Theories}}
\author{Krzysztof Sienicki\thanks{Chair of Theoretical Physics of Naturally Intelligent Systems, Lipowa 2/Topolowa 19, 05-807 Podkowa Leśna, Poland, EU.}}
\date{\today}

\begin{document}

\maketitle

\begin{abstract}
This paper explores the intersection of identity, individuality, and reality through competing frameworks, including classical metaphysics, quantum mechanics, and computational theories. Traditional metaphysical notions of fixed identity are challenged by advancements in cloning, teletransportation, and digital replication, which reveal the fluid and relational nature of individuality. Quantum mechanics further complicates these notions, emphasizing the indistinguishability and contextuality of fundamental particles. Computational approaches, such as the Ruliad and Constructor Theory, offer expansive views of emergent realities but often lack practical constraints for observer relevance. Algorithmic idealism is introduced as a unifying framework, proposing that reality is an emergent construct governed by computational rules prioritizing coherence, sufficiency, and observer-dependent experiences. By redefining identity as an informational construct and reality as a process shaped by algorithmic transitions, algorithmic idealism resolves foundational paradoxes and offers a superior lens for understanding existence in an increasingly digital and interconnected world. The framework bridges gaps between competing theories, providing a coherent and pragmatic model for addressing ethical, metaphysical, and technological challenges.
\end{abstract}

\maketitle

\maketitle

\tableofcontents
\begin{abstract}
This paper explores the intersection of identity, individuality, and reality through competing frameworks, including classical metaphysics, quantum mechanics, and computational theories. Traditional metaphysical notions of fixed identity are challenged by advancements in cloning, teletransportation, and digital replication, which reveal the fluid and relational nature of individuality. Quantum mechanics further complicates these notions, emphasizing the indistinguishability and contextuality of fundamental particles. Computational approaches, such as the Ruliad and Constructor Theory, offer expansive views of emergent realities but often lack practical constraints for observer relevance. Algorithmic idealism is introduced as a unifying framework, proposing that reality is an emergent construct governed by computational rules prioritizing coherence, sufficiency, and observer-dependent experiences. By redefining identity as an informational construct and reality as a process shaped by algorithmic transitions, algorithmic idealism resolves foundational paradoxes and offers a superior lens for understanding existence in an increasingly digital and interconnected world. The framework bridges gaps between competing theories, providing a coherent and pragmatic model for addressing ethical, metaphysical, and technological challenges.

\end{abstract}
\section{Summa Technologiae}

Stanisław Lem (1921-2006)  was a prolific writer and polymath.\footnote{https://culture.pl/en/article/13-things-lem-predicted-about-the-future-we-live-in} He is regarded as one of the most significant science fiction authors in history and one of Poland's most influential literary figures. In 1974, Philip K. Dick, an American science fiction writer, wrote a letter to the FBI alleging that Stanisław Lem was a Polish communist provocateur using a fictitious name, purportedly leading "a faceless group in Kraków, Poland" and producing a prolific and intellectually diverse body of work aimed at manipulating public opinion.\cite{sienicki2016captain}

In Chapter 6 of his philosophical treatise \textit{Summa Technologiae}\footnote{\textit{Summa Technologiae} is one of Stanisław Lem's most influential works, first published in Polish in 1964. The title, a nod to Thomas Aquinas's \textit{Summa Theologiae}, reflects Lem's ambitious goal of providing a comprehensive philosophical exploration of the implications of technology and its impact on humanity, culture, and the future. }, first published in Polish in 1964, Stanisław Lem provides a profound and thought-provoking reinterpretation of individuality in the context of technological progress. He critiques the traditional view of individuality as a fixed and unchanging essence, proposing instead that it is a fluid construct influenced by both environmental factors and technological developments. Lem contends that individuality should not be regarded as an absolute category, emphasizing that:

\begin{quote}
Individuality is not an absolute category; it is a variable construct shaped by the environment and technology.\footnote{In the original: "Indywidualność nie jest kategorią absolutną; jest tworem zmiennym, kształtowanym przez środowisko i technologię."}
\end{quote}

This perspective frames individuality as malleable and subject to continuous transformation as new technologies emerge.

One of Lem’s key insights is the paradox of identity raised by technologies such as cloning and replication. He imagines a future where it is possible to duplicate not only the physical body but also the consciousness, thoughts, and memories of an individual. This scenario forces us to ask what truly distinguishes an individual from their clone. Lem writes:

\begin{quote}
If memory and consciousness can be duplicated, then what distinguishes the individual from their copy?\footnote{In the original "Jeżeli pamięć i świadomość mogą być powielane, to co wówczas odróżnia jednostkę od jej kopii?"}
\end{quote}

In this vision, individuality becomes fluid and potentially distributed across multiple entities, undermining the traditional understanding of the self as singular and indivisible.

Lem further distinguishes between biological and cybernetic individuality. Biological individuality, constrained by the physical and genetic limitations of the human body, contrasts with cybernetic individuality, which transcends these boundaries through technological means. He envisions cybernetic systems capable of creating personalities with traits that biological humans cannot possess, stating:
\begin{quote}
Cybernetic systems may produce personalities with traits inaccessible to biological humans.\footnote{In the original "Cybernetyczne systemy mogą wytwarzać osobowości o cechach niedostępnych dla biologicznego człowieka."}  
\end{quote}
This raises questions about what it means to be an individual and whether artificial systems can be considered as such.

A central theme in Lem’s vision is the \textit{relational} and \textit{contextual} nature of individuality. He suggests that as technological systems integrate individuals into networks, traditional boundaries of identity will dissolve. Lem envisions a future where individuals may exist as multi-person networked structures, sharing consciousness and experience: 

\begin{quote}
In the future, individuals may exist as multiperson networked structures, sharing consciousness and experience.\footnote{In the original "W przyszłości jednostki mogą istnieć jako wieloosobowe struktury sieciowe, dzieląc się świadomością i doświadczeniem."}
\end{quote}
\begin{quote}
    
\end{quote}
This perspective moves away from an essentialist view of the self and embraces a more fluid, interconnected understanding of identity.

Ethical concerns are also central to Lem’s discussion. He warns that technologies capable of shaping individuality could lead to the erosion of autonomy, writing:

\begin{quote}
Technologies that allow for the shaping of personality may lead to the erosion of individual autonomy.\footnote{In the original "Technologie, które pozwalają na kształtowanie osobowości, mogą prowadzić do erozji autonomii jednostki."} 
\end{quote}
This insight anticipates modern debates on the influence of artificial intelligence, genetic engineering, and surveillance technologies on human agency. For Lem, the programmability of individuality poses profound moral dilemmas, questioning whether free will and personal responsibility can endure in a world where the self is increasingly shaped by external forces.

Lem’s vision of individuality in \textit{Summa Technologiae} is both provocative and prescient. By redefining individuality as a flexible and context-dependent construct, he forces us to reconsider the boundaries of the self in the face of transformative technologies. His reflections challenge essentialist views of identity, suggesting that individuality is not a fixed property but a dynamic process shaped by interactions with technology and the environment. As Lem aptly concludes, 

\begin{quote}
The future of humanity will be defined by technologies that not only change the world but also shape who we are.\footnote{In the original "Przyszłość człowieka będzie definiowana przez technologie, które nie tylko zmieniają świat, ale także kształtują to, kim jesteśmy."}
\end{quote}
His work invites us to confront the ethical, philosophical, and existential challenges posed by the technological reshaping of human identity.

\section{Reasons and Persons}

In Chapter 10 of \textit{Reasons and Persons}, Derek Parfit \cite{parfit1987reasons} fundamentally challenges the traditional understanding of personal identity, presenting a groundbreaking view that reshapes how we think about survival, moral responsibility, and ethical decision-making. Parfit rejects the notion that personal identity is a fixed metaphysical concept that must persist for a person to exist over time. Instead, he argues that what truly matters is psychological continuity and connectedness—relationships between mental states like memories, beliefs, and personality traits—rather than a strict, all-or-nothing conception of identity.

Parfit critiques the traditional view of personal identity, which assumes that a person must remain numerically the same to maintain their existence. This view treats identity as binary: you are either the same person or not. Parfit disputes this framework, proposing that identity is not an absolute fact but a matter of degree. In other words, the connections between psychological states are what define the self, and these connections can exist to varying extents, without requiring the preservation of a single, unified identity.

To illustrate this, Parfit employs thought experiments that challenge the necessity of a strict identity. One of his most famous scenarios involves teletransportation. Imagine a person stepping into a machine that destroys their body while creating a perfect replica elsewhere, complete with the same memories, beliefs, and personality. Parfit asks whether the replica is the same person as the original. While the replica may have all the attributes of the original, the destruction of the original body raises questions about whether identity has been preserved. Parfit suggests that the replica’s psychological continuity and connectedness to the original person matter more than the absence of strict numerical identity.

Another thought experiment involves dividing a person’s brain and transplanting each half into two different bodies. If both resulting individuals retain the original’s memories and personality traits, Parfit argues that they are psychologically continuous with the original person. Yet, traditional notions of identity falter in this scenario because identity demands a singular successor. Parfit concludes that strict identity is not necessary for survival; what matters is the psychological connectedness shared by the resulting individuals.

From these examples, Parfit advances his claim that personal identity is not what truly matters. He argues that survival, understood as the preservation of psychological connections, is more important than maintaining a single, unified identity. This view has profound implications for how we approach ethical and practical questions. For example, if identity is not central to survival, the fear of death might diminish. Similarly, moral responsibility could extend to individuals who share significant psychological continuity, even if strict identity is absent.

Parfit’s rejection of identity as a "further fact" also reshapes our understanding of ethical decision-making. If psychological continuity matters more than identity, ethical considerations shift toward preserving connections between mental states rather than ensuring the persistence of a singular self. This has practical implications for how we think about cloning, brain transplants, or other scenarios that challenge traditional boundaries of selfhood.

Derek Parfit’s teletransportation paradox plays a foundational role in the philosophical arguments of Algorithmic Idealism, particularly in reshaping our understanding of identity, continuity, and subjective experience. Parfit’s thought experiment presents a scenario where a person undergoes teletransportation: their original body is destroyed, and an exact copy, complete with all memories and personality traits, is recreated elsewhere. This paradox challenges traditional views of personal identity, especially those grounded in physical continuity or the persistence of a singular body. In Algorithmic Idealism, Parfit’s paradox is employed to illuminate the limitations of physicalist and materialist frameworks, ultimately supporting a shift to a radically informational and algorithmic approach to understanding identity and experience.

Parfit’s paradox raises a central question: if the teletransportation process creates an exact duplicate, is the individual who emerges the same as the one who entered? Traditional frameworks often falter when addressing such scenarios. They attempt to tether identity to material continuity, asserting that identity relies on the uninterrupted existence of the physical body or brain. However, in the context of teletransportation or duplication, these assumptions collapse. The original body is destroyed, and yet the recreated individual is, in every observable way, identical to the original. Parfit’s thought experiment exposes the insufficiency of conventional metaphysical theories to grapple with identity in non-traditional contexts, creating fertile ground for the algorithmic reconceptualization proposed by Algorithmic Idealism.

Algorithmic Idealism offers a transformative perspective by rejecting the physicalist reliance on material continuity. Instead, it reframes identity as an informational construct defined by self-states—abstract patterns that encode the entirety of an agent’s experience at a given moment. From this perspective, the question of whether the teletransported individual is “the same” becomes irrelevant. If the informational structure of the recreated self-state is identical to the original, then the identity of the individual is preserved. Algorithmic Idealism shifts the focus from material persistence to the coherence and continuity of informational patterns, resolving the paradox by treating all instances of a self-state as equivalent realizations of the same identity.

This reframing is not merely a response to metaphysical challenges but also an epistemological shift. Parfit’s paradox highlights the inadequacy of third-person, externalist perspectives in addressing questions of first-person identity. The traditional focus on “what exists in the world” fails to provide meaningful answers to subjective questions like “What will I experience next?” Algorithmic Idealism embraces this subjective perspective, positing that reality is best understood as a sequence of transitions between self-states. These transitions are governed by algorithmic probabilities, providing a mathematically rigorous framework for predicting what an agent should expect to experience next, without reference to an external world.

Parfit’s paradox thus supports Algorithmic Idealism’s broader critique of physicalism. In scenarios involving duplication, traditional physical theories can describe the external facts—how many copies exist, where they are located—but they cannot answer the essential first-person question: “Which one am I?” This gap underscores the need for a framework like Algorithmic Idealism, which prioritizes subjective experience over external description. By focusing on the informational structure of self-states, the framework bypasses the metaphysical confusion of “original” versus “copy” and instead grounds identity in the coherence of informational transitions.

Furthermore, Algorithmic Idealism’s resolution of Parfit’s paradox has profound implications for ethics, metaphysics, and even technology. In ethical terms, it suggests that concerns about the destruction or duplication of physical bodies are secondary to the preservation of informational structures. This perspective redefines debates around brain uploading, cloning, and digital resurrection, emphasizing the continuity of self-states rather than the material substrate in which they are realized. Metaphysically, it dissolves the distinction between physical and simulated realities. A teletransported individual, a digital simulation, and a biological body are all equivalent as long as the informational coherence of the self-state is maintained. This insight not only resolves Parfit’s paradox but also reframes broader philosophical questions about the nature of existence and identity in an increasingly digital world.

Derek Parfit’s teletransportation paradox serves as a critical foundation for the arguments of Algorithmic Idealism. By exposing the limitations of traditional, materialist notions of identity, it supports the shift to an informational framework that defines identity through self-states and their algorithmic transitions. Algorithmic Idealism resolves the paradox by treating all instances of a self-state as equivalent, eliminating the need for distinctions between originals and copies. This approach redefines identity as an emergent property of informational coherence, offering a profound and flexible framework for understanding personal identity in the digital and quantum age. Parfit’s paradox, far from being an abstract thought experiment, becomes a cornerstone in the rethinking of reality as an interplay of informational patterns, guiding us toward a deeper understanding of identity and experience.

\section{Identity and Individuality}

The concept of identity and individuality is profoundly challenged both in the realm of quantum mechanics and in Derek Parfit’s philosophical exploration of personal identity through the teletransportation paradox. Both frameworks question traditional notions of what it means for something or someone to maintain identity over time or across transformations, ultimately suggesting that identity might not be a fundamental feature of reality, but rather a functional or relational construct.

In quantum mechanics, the indistinguishability of identical particles such as electrons raises questions about individuality and identity. Unlike macroscopic objects, quantum particles cannot be assigned unique labels, even in principle. This property directly challenges Leibniz’s Principle of the Identity of Indiscernibles, which states that no two distinct objects can have all their properties in common. In quantum systems, particles exist as part of a larger, indistinguishable ensemble, where their individuality is subsumed into the statistical description provided by quantum states. For instance, the wavefunction of a pair of identical fermions is antisymmetric under particle exchange, making it impossible to treat these particles as independently identifiable entities. This collapse of individuality on the quantum level forces a reevaluation of what it means to be “identical” and whether such entities can possess an identity distinct from their role in the system.

Parfit’s teletransportation paradox raises similar challenges about identity, but in the context of personal identity and human continuity. The paradox describes a scenario in which a person’s body is destroyed and then perfectly recreated elsewhere, with the recreated individual retaining all memories, traits, and personality of the original. The central question becomes whether the recreated individual is the same person as the original. Traditional views of identity rooted in physical continuity or the persistence of a singular body falter in this scenario, as the original body no longer exists. Parfit’s paradox forces us to reconsider whether identity is tied to physical continuity or if it can instead be understood as a functional or informational construct.

The parallels between these two domains are striking. In both quantum mechanics and the teletransportation paradox, identity is revealed to be context-dependent rather than absolute. Quantum mechanics shows that identity and individuality at the particle level may not exist in any fundamental sense, as indistinguishable particles do not have unique, intrinsic identifiers. Similarly, in Parfit’s paradox, the distinction between the original and the recreated individual becomes meaningless if identity is understood not as a metaphysical essence, but as a continuity of properties or patterns. If the recreated individual in Parfit’s scenario perfectly matches the original in every functional sense, it challenges the necessity of a singular, unique “self” that persists unbroken through time.

Both frameworks also emphasize the limitations of traditional metaphysical assumptions in addressing questions of identity. In the quantum realm, the inability to distinguish identical particles undermines the classical notion of individuality, suggesting instead that identity might emerge only relationally or contextually within systems. Likewise, in Parfit’s paradox, the inability to privilege the “original” over the “copy” disrupts the classical view of a unique, indivisible self. In both cases, the focus shifts from intrinsic identity to the relational or functional properties that define how entities behave and interact within a given framework.

These insights have broader implications for our understanding of identity, individuality, and existence. The quantum mechanical perspective implies that identity may not be a fundamental feature of nature, but rather an emergent property of systems. This view aligns with Parfit’s suggestion that personal identity might not be tied to physical or metaphysical continuity, but instead to informational patterns or functional equivalence. Together, these perspectives challenge us to rethink identity not as something static or intrinsic, but as something dynamic and context-dependent.

The interplay between quantum mechanics and Parfit’s teletransportation paradox reveals profound parallels in how identity and individuality are conceptualized. Both challenge the classical notions of intrinsic identity, suggesting instead that identity may be an emergent, relational property tied to patterns and functions rather than physical continuity or metaphysical essence. These frameworks push us toward a more flexible and nuanced understanding of identity, one that accommodates the complexities of both the quantum world and the human experience. As our understanding of these concepts evolves, it opens new avenues for exploring identity in fields as diverse as metaphysics, ethics, and the philosophy of mind.

\section{Distinguishability and Accessible Information}

The exploration of quantum mechanics, particularly the concepts of distinguishability and accessible information, reveals profound insights into the nature of reality and its relationship with observation and knowledge. \cite{bigaj2022identity, zurek2003decoherence} These ideas suggest that reality, far from being an objective and independent entity, is shaped by the interactions and limitations inherent in the systems we use to observe and understand it. Such a perspective resonates with a broader philosophical viewpoint that reality itself might be fundamentally emergent, constructed through processes governed by certain rules or constraints. \cite{rovelli2018physics, wallace2012emergent}

One of the central ideas in quantum mechanics is the limitation on our ability to distinguish between different states of a system . \cite{nielsen2010quantum, hardy2001quantum} This limitation is not a flaw but a feature of how reality operates at its most fundamental level. It implies that our access to information is inherently constrained, reflecting a system that provides only what is necessary for interaction and coherence. These constraints suggest that the universe might operate according to rules that prioritize sufficiency over completeness, ensuring that reality remains structured and comprehensible rather than overwhelming or chaotic. \cite{spekkens2007evidence}

The notion that quantum states represent not objective truths but informational constructs is another important insight. \cite{brukner2014quantum} In this view, quantum states describe what we know—or can know—about a system, rather than what the system "is" in an absolute sense. This shifts the focus from a reality that exists independently of observation to one that is shaped by the interaction between the observer and the observed. It underscores the idea that knowledge and reality are interconnected, and what we perceive as reality is deeply tied to the frameworks and processes through which we interact with the world. \cite{zurek2003decoherence, spekkens2007evidence}

A particularly compelling aspect of quantum mechanics is the relationship between information and disturbance. The act of observation or measurement inevitably alters the system being observed, creating a dynamic interplay between the observer and the system. This dynamic suggests that reality is not static or fixed but emerges through interactions that reshape it. Such a perspective highlights the active role of the observer in the construction of reality, where the process of engaging with the world fundamentally influences what is observed. \cite{rovelli2018physics}

Limits on accessible information, such as those described by quantum mechanics, further emphasize the idea of a structured and emergent reality. \cite{nielsen2010quantum, hardy2001quantum} These bounds imply that the universe is not infinitely knowable; instead, it offers a finite and bounded framework within which interactions take place. This boundedness ensures that reality remains consistent and manageable, allowing for coherent interactions and experiences. It suggests a reality that is shaped by rules designed to balance complexity with comprehensibility. \cite{zurek2003decoherence}

The inability to replicate or broadcast quantum states, as highlighted in quantum mechanics, reinforces this view of reality as contextual and localized. \cite{nielsen2010quantum} Such constraints prevent universal duplication and ensure that reality maintains its coherence and uniqueness. This highlights the importance of local interactions in shaping the structures and experiences that emerge, suggesting that reality is not a universal, one-size-fits-all construct but a tapestry woven from countless individual interactions. \cite{spekkens2007evidence}

Taken together, these ideas challenge traditional notions of an objective, observer-independent reality. Instead, they suggest a view of reality as emergent, dynamic, and shaped by interaction. The rules or constraints governing these interactions ensure that reality remains structured and meaningful, offering just enough information and coherence to sustain interaction and understanding. \cite{brukner2014quantum, wallace2012emergent} This perspective invites us to rethink our relationship with the world, seeing reality not as a pre-existing entity to be uncovered but as a process to be engaged with and understood through interaction. It opens up a view of the universe as a place of ongoing creation, where reality and knowledge continuously emerge through the interplay of observation, interaction, and the underlying rules that govern them.

Quantum mechanics and the principles of algorithmic idealism converge on this understanding, providing complementary perspectives on the nature of reality and its dynamic emergence. Algorithmic idealism posits that reality is not static or predetermined but emerges through processes governed by computational rules. \cite{hardy2001quantum} This philosophical viewpoint mirrors quantum theory's insights into the observer-dependent and constrained nature of reality. The limitations on distinguishability, the interplay between information and disturbance, and the bounded nature of accessible information all align with the idea that reality is structured and governed by principles that balance complexity and coherence.

Quantum states, as informational constructs, embody the algorithmic idealist notion that reality is not an intrinsic truth but a product of interaction. \cite{brukner2014quantum, spekkens2007evidence} The dynamic interplay between observer and system in quantum measurement reflects the computational process by which reality is shaped. This perspective emphasizes the active role of the observer in constructing knowledge and highlights the emergent nature of reality as a continuous process of engagement and interaction.

Similarly, the boundedness of reality in quantum mechanics—such as limits imposed by the Holevo bound—reflects the principle of sufficiency over completeness. \cite{nielsen2010quantum} In algorithmic idealism, reality is seen as providing just enough information to enable meaningful interaction and coherence. This balance ensures that reality remains accessible and intelligible without overwhelming complexity, reinforcing the idea that reality is a structured construct governed by rules.

The localized and contextual nature of reality, as emphasized by the no-broadcasting theorem, also resonates with algorithmic idealism. \cite{spekkens2007evidence, nielsen2010quantum} This theorem illustrates that reality is not universal or monolithic but emerges through unique, localized interactions. The preservation of coherence and the prevention of universal replication ensure that reality maintains its distinct and emergent character, further aligning with the idealist view of a dynamically constructed universe.

Quantum mechanics and algorithmic idealism share a deep philosophical connection. Both frameworks challenge traditional notions of objective reality, emphasizing instead a dynamic, emergent process shaped by interaction and governed by rules or constraints. The insights of quantum mechanics, from the limits on distinguishability to the interplay of information and disturbance, reflect the principles of algorithmic idealism, where reality is constructed through processes that balance complexity, coherence, and accessibility. \cite{rovelli2018physics, wallace2012emergent} Together, these perspectives invite us to view reality not as something to be discovered but as something that emerges through interaction, observation, and the fundamental rules that define existence. This synthesis deepens our understanding of the nature of reality and highlights the interconnectedness of knowledge, interaction, and the fundamental structure of the universe.

\section{No-cloning and Algorithmic No-cloning}

The no-cloning theorem is a fundamental result in quantum mechanics that prohibits the perfect duplication of arbitrary quantum states. This prohibition stems from the linearity of quantum mechanics and the superposition principle, which make it impossible to replicate unknown quantum states without violating the fundamental properties of quantum systems. \cite{nielsen2010quantum} Rather than being a technical limitation, this theorem highlights the contextual and localized nature of quantum information, ensuring the coherence and uniqueness of quantum systems. \cite{zurek2003decoherence}

The no-cloning theorem has far-reaching implications, particularly in the domain of quantum information theory and quantum cryptography. For example, it ensures the security of quantum key distribution protocols, where any attempt to eavesdrop disturbs the transmitted quantum states, thus revealing the intrusion. \cite{spekkens2007evidence} Beyond quantum mechanics, this idea resonates with broader computational and philosophical concepts, particularly algorithmic idealism.

Algorithmic no-cloning extends the principles of no-cloning to systems governed by computational rules. In this perspective, the inability to replicate complex systems or processes arises from inherent computational constraints. This aligns with the philosophy of algorithmic idealism, which proposes that reality emerges from algorithmic processes governed by rules balancing complexity and coherence. \cite{hardy2001quantum, brukner2014quantum} In this framework, reality is not an objective, static entity but an emergent construct shaped by the interactions and constraints defining it.

The philosophical insights of algorithmic idealism echo quantum mechanics' emphasis on observer-dependence and information constraints. The no-cloning theorem, when viewed through this lens, underscores the interconnectedness of observation, knowledge, and the nature of reality. Reality is thus shaped by processes that maintain coherence and uniqueness while preventing redundancy or overwhelming complexity. \cite{wallace2012emergent, rovelli2018physics} The localized and contextual nature of quantum information, as highlighted by the no-cloning theorem and the no-broadcasting theorem, reinforces this perspective, suggesting that reality emerges through interactions that preserve its structured and bounded nature.\cite{zurek2003decoherence}

The interplay between information and disturbance in quantum mechanics provides further support for the emergent and algorithmically governed nature of reality. Observation inevitably alters the system being observed, creating a dynamic interplay that reflects the process of reality’s construction. This dynamic suggests that reality is not static but emerges through interactions and rules ensuring that it remains comprehensible and meaningful. \cite{brukner2014quantum}

The synthesis of the no-cloning theorem and algorithmic idealism challenges traditional notions of objective, observer-independent reality. Instead, it invites us to see reality as a process shaped by computational principles and observer interactions. These principles ensure that reality provides sufficient information for meaningful interaction and coherence, avoiding overwhelming complexity or ambiguity. By integrating these perspectives, we gain a deeper understanding of reality as a structured and emergent phenomenon governed by the interplay of observation, interaction, and computational rules.

\section{Quantum Darwinism and Algorithmic Idealism}

The theory of quantum Darwinism offers a compelling explanation for how the classical world emerges from quantum systems. It builds on the concepts of decoherence and information proliferation, proposing that the redundancy of information about certain states enables the appearance of objective, classical reality. This idea aligns closely with algorithmic idealism, which views reality as an emergent construct shaped by computational processes and observer interactions. Together, quantum Darwinism and algorithmic idealism challenge traditional notions of reality, offering a perspective where information, observation, and interaction are central to the construction of the universe.

Quantum Darwinism, proposed by Wojciech Zurek, extends the principles of decoherence to explain why specific quantum states become classical. In this framework, the environment plays a crucial role as a medium that encodes and disseminates information about certain "pointer states." These states are robust against decoherence and can be redundantly recorded in the environment, allowing multiple observers to independently access the same information. \cite{zurek2009quantum} This reedundancy creates the illusion of objective reality, as the environment effectively "selects" classical states that are stable and reproducible. \cite{zurek2003decoherence}

Algorithmic idealism complements quantum Darwinism by suggesting that this selection process is governed by computational rules that balance complexity, coherence, and accessibility. From the algorithmic idealist perspective, the redundancy of classical states in quantum Darwinism reflects an underlying principle of sufficiency: the universe encodes just enough information to ensure meaningful interactions without overwhelming complexity. This balance ensures that reality remains comprehensible and structured, aligning with the idea of emergent reality shaped by computational constraints. \cite{brukner2014quantum, hardy2001quantum}

The interplay between observation and the emergence of classical reality further underscores the connection between quantum Darwinism and algorithmic idealism. In both frameworks, reality is not a static, pre-existing entity but a dynamic construct shaped by interactions and the dissemination of information. The redundancy of classical states in quantum Darwinism parallels the algorithmic idealist notion that reality is structured to prioritize coherence and interaction over completeness. \cite{rovelli2018physics} This perspective challenges traditional metaphysical views, proposing instead that reality is a process of continuous construction governed by principles of interaction and computation.

Quantum Darwinism and algorithmic idealism also converge on the idea of bounded reality. The redundancy of classical information is limited by environmental constraints, ensuring that the emergence of classical states does not lead to infinite complexity. Similarly, algorithmic idealism posits that reality is bounded by computational rules that prevent overwhelming detail or redundancy. These constraints reflect a universe designed for interaction and comprehension, where reality is shaped by principles that prioritize accessibility and coherence. \cite{zurek2009quantum, wallace2012emergent}

Quantum Darwinism provides a physical mechanism for the emergence of classical states, while algorithmic idealism offers a philosophical interpretation of these processes as governed by computational principles. Together, they emphasize the role of information, observation, and interaction in shaping the universe, inviting us to see reality as an emergent, dynamic construct rather than a static, objective entity.

\section{Quantum Darwinism, Algorithmic Idealism, and Boltzmann Brains
}
The emergence of reality, the nature of observation, and the constraints governing existence are central themes explored in Quantum Darwinism, Algorithmic Idealism, and the paradoxical concept of the Boltzmann Brain. These ideas, though arising from distinct disciplines—quantum mechanics, computational philosophy, and cosmology—intersect on critical questions about the structure of reality, the role of observers, and the mechanisms that ensure coherence over randomness. Together, they reveal complementary and contrasting perspectives on how reality and observers emerge.

A Boltzmann Brain is a hypothetical self-aware observer that spontaneously forms due to random fluctuations in a thermodynamic system. This concept emerges from cosmological models that suggest improbable configurations of matter can arise in infinite or sufficiently vast universes. Boltzmann Brains highlight the tension between structured, emergent reality and chaotic randomness. The randomness underpinning their formation contrasts sharply with the mechanisms described by Quantum Darwinism, which explains the emergence of classical reality through environmental decoherence. In Quantum Darwinism, robust "pointer states" of quantum systems are redundantly encoded in the environment, enabling multiple observers to access the same information and creating a consistent, shared reality. This structured process ensures the stability and coherence of classical reality, in stark opposition to the isolated and random formation of Boltzmann Brains. \cite{zurek2009quantum}

Quantum Darwinism aligns closely with the principles of Algorithmic Idealism, which posits that reality emerges through computational processes governed by rules that prioritize sufficiency, coherence, and simplicity. In this framework, reality is not a chaotic, random entity but an emergent structure shaped by interactions and constraints. Algorithmic Idealism suggests that the improbability of Boltzmann Brains reflects the underlying computational design of the universe, which suppresses randomness and favors structured, meaningful interactions. Unlike Boltzmann Brains, which exist as isolated phenomena in low-probability configurations, observers in Algorithmic Idealism emerge from systematic processes that balance complexity with coherence. \cite{brukner2014quantum, hardy2001quantum}

The no-cloning theorem and no-broadcasting theorem in quantum mechanics further reinforce the structured nature of emergent reality. \cite{horodecki2006quantumness} These theorems impose constraints on the replication and distribution of quantum information, ensuring that observers and interactions remain localized and contextual. These rules prevent the unbounded duplication of quantum states and maintain coherence within the quantum framework. Boltzmann Brains, however, challenge this structure by existing outside the bounds of information constraints. Their formation is not tied to coherent interactions or redundancy but arises from random fluctuations, which lack the systematic processes emphasized in Quantum Darwinism and Algorithmic Idealism. \cite{zurek2003decoherence}

Both Quantum Darwinism and Algorithmic Idealism reject the chaotic emergence of observers suggested by the Boltzmann Brain concept. Instead, they describe reality as emergent, structured, and shaped by redundant processes and computational rules. Quantum Darwinism provides a mechanism for the appearance of objective classical reality through the proliferation of pointer states in the environment, enabling shared experiences among observers. Algorithmic Idealism further asserts that reality operates within a computational framework that ensures coherence and sufficiency, avoiding overwhelming complexity or redundancy. This structured emergence starkly contrasts with the randomness of Boltzmann Brain scenarios, which represent a breakdown of coherence and interaction . \cite{wallace2012emergent}

The improbability of Boltzmann Brains in a structured universe governed by computational rules highlights the role of constraints and mechanisms that suppress chaos in favor of order. Boltzmann Brains emerge in cosmological models where probability favors random fluctuations over structured systems, particularly in infinite universes. However, both Quantum Darwinism and Algorithmic Idealism suggest that emergent reality is not solely a product of statistical likelihood but arises from mechanisms that constrain randomness and favor coherence. Quantum Darwinism achieves this through the selection of stable, classical states via environmental decoherence, while Algorithmic Idealism enforces computational principles that guide the emergence of structured observers. \cite{carroll2017boltzmann, albrecht2004boltzmann}

Ultimately, the relationship between Boltzmann Brains, Quantum Darwinism, and Algorithmic Idealism highlights contrasting perspectives on how reality and observers emerge. Boltzmann Brains represent chaotic, random formations, whereas Quantum Darwinism and Algorithmic Idealism emphasize structured, rule-based processes that prioritize coherence, interaction, and sufficiency. Together, these frameworks deepen our understanding of reality as a dynamic and emergent construct governed by computational and informational constraints, rejecting the chaotic randomness of Boltzmann Brains in favor of structured, meaningful existence.

Here, for the second time, we briefly reference Stanisław Lem's work. Stanisław Lem’s \textit{Solaris} published in 1961, offers a compelling metaphor for exploring emergent reality, observer-dependence, and the role of information. When juxtaposed with the Boltzmann Brain paradox and Algorithmic Idealism, it highlights contrasting perspectives on the structured versus random emergence of observers and realities. \cite{jankovic2022gaia}

\textit{Solaris} portrays a planet-wide, hyper-integrated system capable of emergent intelligence and interaction. This stands in sharp contrast to Boltzmann Brains—hypothetical self-aware entities arising randomly in infinite systems. While Boltzmann Brains represent chaotic, context-free phenomena, \textit{Solaris} embodies structured, feedback-driven emergence, aligning with Algorithmic Idealism’s principle that reality is shaped by computational rules prioritizing coherence and relevance.\footnote{\textit{Solaris} (2002), an American film directed by Steven Soderbergh and produced by James Cameron, stars George Clooney. The film focuses on human relationships, largely omitting Lem's scientific and philosophical themes once again.}

Algorithmic Idealism posits that reality is a bounded computational construct, optimized for observer interaction and sufficiency. Similarly, \textit{Solaris} reflects the integration of information into a functional and coherent whole. The ocean’s ability to interact with humans while maintaining planetary-level homeostasis mirrors the structured realities described by Algorithmic Idealism. In contrast, Boltzmann Brains lack this coherence, underscoring the importance of constraints in creating meaningful and intelligible experiences.

Both Lem’s vision and Algorithmic Idealism reject naive realism, suggesting that reality is not a fixed, objective entity but an emergent phenomenon shaped by interaction and computation. Together, \textit{Solaris}, Boltzmann Brains, and Algorithmic Idealism explore the boundaries of randomness and structure, illustrating how constraints and feedback mechanisms are crucial for creating intelligible, observer-relevant realities.

\section{The Ruliad and Algorithmic Idealism}

The Ruliad, introduced by Stephen Wolfram, and Algorithmic Idealism, proposed by Markus Müller, both explore the emergence of reality from computational processes. While the Ruliad represents the totality of all possible computational rules and their consequences, Algorithmic Idealism narrows its focus to the subset of these processes that generate structured, coherent, and observer-relevant realities. Together, these frameworks offer complementary perspectives on the computational foundations of reality, highlighting the interplay between infinite possibilities and bounded observer experiences.

The Ruliad is an infinite and entangled structure that encompasses every conceivable computational rule and its outcomes. Within this framework, all possible transformations and interactions coexist, making it a "universal object" that encodes every potential reality. Observers, constrained by their computational limitations and histories, perceive only a specific "thread" of the Ruliad—one that aligns with their ability to process and interpret information. \cite{wolfram2021ruliad} This suggests that reality, as experienced by any observer, is deeply tied to their computational perspective and capabilities. Instead of accessing the entirety of the Ruliad, observers interact with only the parts relevant and accessible to their cognitive and physical constraints.

Algorithmic Idealism complements this by asserting that the reality experienced by observers is not just a random slice of the Ruliad but a construct shaped by computational rules that prioritize coherence, sufficiency, and interaction. Müller argues that reality emerges as a structured and bounded construct designed to provide "just enough" information for observers to engage meaningfully with it. This principle of algorithmic sufficiency ensures that observers encounter a comprehensible and navigable reality, avoiding the overwhelming complexity of the Ruliad’s vastness. \cite{muller2021algorithmic}

Both frameworks emphasize observer-dependence in their treatment of reality. The Ruliad describes how the specific slice of reality perceived by an observer is determined by their computational constraints and interactions with the underlying processes. Similarly, Algorithmic Idealism views reality as fundamentally epistemic, arising from the interplay between computational rules and the observer’s cognitive framework. Both perspectives reject the notion of a single, objective reality, emphasizing instead a dynamic, observer-centered view.

The Ruliad's vastness highlights the sheer scope of computational possibilities. It encompasses every conceivable rule, law, and transformation, providing a theoretical foundation for exploring all possible realities. However, this infinite complexity also presents challenges, as most regions of the Ruliad are irrelevant or inaccessible to observers. Algorithmic Idealism addresses this by narrowing the focus to computational rules that produce structured, meaningful realities tailored to observers’ epistemic and computational limits. This pragmatic filtering of possibilities reflects the sufficiency principle, where reality is constructed to balance complexity with intelligibility.

Computational irreducibility is a key concept that connects the Ruliad and Algorithmic Idealism. In the Ruliad, many processes are irreducible, meaning their outcomes cannot be predicted without explicitly running the computation. This irreducibility highlights the inherent complexity of computational processes. Algorithmic Idealism embraces this principle while emphasizing that the complexity of reality is bounded and structured to remain intelligible to observers. By focusing on the accessible and observer-relevant regions of the Ruliad, Algorithmic Idealism ensures that irreducibility does not overwhelm the observer's capacity to navigate and understand reality. \cite{wolfram2002nks, muller2021algorithmic}

Another significant intersection between the Ruliad and Algorithmic Idealism lies in their treatment of universality. The Ruliad encompasses all possible rules and transformations, serving as a universal framework for describing reality. Algorithmic Idealism builds on this idea, asserting that while universality is a foundational feature of reality, only a subset of universal rules contributes to the structured, observer-relevant realities we experience. This refinement aligns with the focus on coherence and sufficiency, ensuring that only the computational processes capable of generating consistent and meaningful experiences are relevant.

The relationship between the Ruliad and Algorithmic Idealism illustrates the tension between infinite computational possibilities and finite observer capacities. While the Ruliad represents the vastness of all possible rules, Algorithmic Idealism imposes constraints that filter this infinite potential into manageable and meaningful realities. These constraints ensure that reality provides a coherent and structured experience, balancing complexity with the observer's computational limitations. Together, these frameworks offer a unified perspective on the emergence of reality, emphasizing the central role of computation, constraints, and observer interactions.

\section{Constructor Theory and Algorithmic Idealism}

Constructor Theory, introduced by David Deutsch and Chiara Marletto, and algorithmic idealism, advanced by Markus Müller, represent innovative approaches to understanding the nature of reality. Both frameworks emphasize the central role of information and computation, but they approach this from different perspectives. Constructor Theory focuses on the constraints governing physical transformations, while Algorithmic Idealism explores the emergent, observer-dependent nature of reality shaped by computational processes. Together, they provide complementary insights into the interplay between physical and epistemic structures.

Constructor Theory reframes the foundations of physics by shifting the focus from dynamical laws to counterfactuals—statements about which physical transformations are possible or impossible. It introduces the concept of a "constructor," an entity capable of performing a specific transformation repeatedly without degrading. By defining physical laws in terms of the constraints that determine what tasks can or cannot be accomplished, Constructor Theory elevates information to a foundational role in understanding reality. For example, the no-cloning theorem in quantum mechanics, which states that arbitrary quantum states cannot be perfectly copied, is recast as a statement about the impossibility of constructing a universal cloning process. \cite{deutsch2015constructor, marletto2015constructor}

Algorithmic idealism, in contrast, focuses on the epistemic and computational emergence of reality. It posits that reality is not a static, objective entity but an emergent phenomenon shaped by computational rules that prioritize coherence, sufficiency, and relevance for observers. This framework emphasizes the boundedness of information, asserting that reality provides "just enough" structure for observers to interact meaningfully with it. Instead of assuming a single, objective reality, Algorithmic Idealism highlights the role of constraints in shaping observer-specific experiences. \cite{muller2021algorithmic, mueller2020law, mueller2024algorithmic}

The relationship between these two frameworks lies in their shared focus on constraints. Constructor Theory examines the physical constraints that determine the feasibility of tasks, emphasizing the role of information in defining these limits. Algorithmic Idealism explores similar constraints from an epistemic perspective, focusing on the computational rules that filter infinite possibilities into structured, intelligible realities. Both reject reductionism, highlighting that reality is better understood through the lens of what is possible and what information is preserved or transformed.

Another key connection is their treatment of information as foundational. Constructor Theory places information at the heart of physical laws, describing the universe in terms of transformations of information. For example, stable classical states emerge because they preserve information in ways that are robust to perturbations. Similarly, Algorithmic Idealism asserts that reality itself is an informational construct, emerging from computational processes that generate coherent and observer-relevant phenomena. This shared focus underscores the primacy of information in shaping both physical and experiential realities.

The concept of universality further unites these frameworks. Constructor Theory introduces the idea of universal constructors, entities capable of performing any physically possible task. This universality mirrors Algorithmic Idealism’s assertion that reality is governed by computational principles that are universal but constrained by the epistemic and computational capacities of observers. While Constructor Theory applies universality to physical systems, Algorithmic Idealism refines this scope, focusing on the subset of computational rules that give rise to structured, observer-relevant realities.

Despite their similarities, the two frameworks diverge in their emphasis. Constructor Theory is explicitly concerned with the principles governing physical systems and the constraints that define transformations. It offers a counterfactual perspective on the laws of physics, focusing on what can and cannot be achieved. Algorithmic Idealism, on the other hand, centers on the epistemic emergence of reality, emphasizing how computational processes shape the experiences of observers. While Constructor Theory provides a framework for describing physical possibilities, Algorithmic Idealism applies these principles to the specific context of how reality is perceived and understood.

Together, constructor theory and algorithmic idealism offer a holistic view of reality that bridges the gap between physical and epistemic perspectives. Constructor theory provides a foundational framework for understanding physical laws in terms of information and counterfactuals, while Algorithmic Idealism explores how these principles manifest in observer-dependent realities. By emphasizing constraints, information, and computation, both frameworks challenge traditional metaphysical assumptions, offering a unified vision of reality as a dynamic interplay between physical possibilities and computational structures.

\section{Summary and Conclusions}

The paper investigates competing theories about identity, individuality, and the nature of reality, including classical metaphysical views, quantum mechanics-informed interpretations, and computational frameworks. Classical metaphysics regards identity as a fixed, intrinsic property tied to physical continuity, but this view is challenged by scenarios like cloning and teletransportation, which reveal its limitations. Quantum mechanics offers a different perspective, emphasizing the indistinguishability and contextual nature of particles, but its focus remains primarily on physical systems. Computational approaches, such as the Ruliad and Constructor Theory, expand these ideas by exploring the infinite potential of emergent realities and the constraints that govern them.

Among these, algorithmic idealism emerges as the most compelling framework. Unlike classical metaphysics, it does not anchor identity to material persistence, and unlike quantum mechanics, it extends beyond physical systems to address the broader informational and experiential structures of reality. Algorithmic idealism frames reality as an emergent construct shaped by computational rules that prioritize coherence, sufficiency, and relevance for observers. This approach not only resolves paradoxes of identity, such as whether a clone or teletransported copy retains individuality, but also redefines reality itself as a dynamic, observer-dependent process.

Algorithmic idealism is superior to other theories because it balances complexity with accessibility. Classical metaphysics and physicalist theories often struggle with explaining identity in non-continuous or non-material contexts, while computational approaches like the Ruliad can overwhelm with infinite possibilities. Algorithmic idealism narrows the focus to meaningful, structured realities that are both comprehensible and relevant to observers, grounded in algorithmic sufficiency rather than exhaustive complexity. This perspective makes it uniquely suited to address the challenges of modern technology, ethics, and metaphysics, offering a practical and philosophically robust framework for understanding individuality and reality in an increasingly computational world.

\end{document}